# Model-based geometrical optimisation and *in vivo* validation of a spatially selective multielectrode cuff array for vagus nerve neuromodulation

Kirill Aristovich[1], Matteo Donega[2], Cathrine Fjordbakk[3], Ilya Tarotin[1], Christopher A. R. Chapman[1], Jaime Viscasillas[3], Thaleia-Rengina Stathopoulou[3], Abbe Crawford[3], Daniel Chew[2], Justin Perkins[3], and David Holder[1]

[1] Medical Physics and Biomedical Engineering, University College London, UK
[2] Neuromodulation, Galvani Bioelectronics, Stevenage, UK
[3] Royal Veterinary College, Hertfordshire, UK

E-mail: k.aristovich@ucl.ac.uk

**Abstract**

*Background*. Neuromodulation by electrical stimulation of the human cervical vagus nerve may be limited by adverse side effects due to stimulation of off-target organs. It may be possible to overcome this by spatially selective stimulation of peripheral nerves. Preliminary studies have shown this is possible using a cylindrical multielectrode human-sized nerve cuff in vagus nerve selective neuromodulation.
*New method*. The model-based optimisation method for multi-electrode geometric design is presented. The method was applied for vagus nerve cuff array and suggested two rings of 14 electrodes, 3 mm apart, with 0.4 mm electrode width and separation and length 0.5-3 mm, with stimulation through a pair in the same radial position on the two rings. The electrodes were fabricated using PDMS-embedded stainless steel foil and PEDOT: pTS coating.
*Results*. In the cervical vagus nerve in anaesthetised sheep, it was possible to selectively reduce the respiratory breath rate (RBR) by 85 ±5% without affecting heart rate, or selectively reduce heart rate (HR) by 20 ± 7% without affecting respiratory rate. The cardiac- and pulmonary-specific sites on the nerve cross-sectional perimeter were localised with a radial separation of 105 ± 5 degrees ($P < 0.01$, $N = 24$ in 12 sheep).
*Conclusions*. Results suggest organotopic or function-specific organisation of neural fibres in the cervical vagus nerve. The optimised electrode array demonstrated selective electrical neuromodulation without adverse side effects. It may be possible to translate this to improved treatment by electrical autonomic neuromodulation for currently intractable conditions.

## 1. Introduction

In recent years, electrical neuromodulation of peripheral nerves to relieve various conditions has received a dramatic increase in interest in the academic and industrial community [1]–[3]. The concept is to place a small cuff device around a nerve to send electrical impulses, generating or blocking action potentials, which are transmitted to the organ of interest and hence modulate the function of the organ. Such devices are currently used in clinics for the management of drug-resistant epilepsy [4], chronic heart failure [5], [6], stroke rehabilitation [7], chronic inflammation [8], and also other disease models [9] have promising results in animals. Due to the relative ease of surgical access, most of the current peripheral nerve interfacing devices are implanted in the neck to interface with the cervical vagus, the main autonomic trunk to all organs of the body [10]. However, there is an ongoing investigation of their use in smaller peripheral nerves such as sacral [11] or tibial [12] nerves for overactive bladder, peroneal nerve for foot drop [13], hypoglossal nerve for obstructive sleep apnoea [14], or occipital nerve for chronic pain management [15].

Although there have been promising results in animal studies, few vagus nerve stimulation devices reach clinical approval. Current autonomic nerve interfaces stimulate the nerve bundle indiscriminately. As the vagus nerve innervates numerous organs, including the heart, lungs, intestines, spleen, and larynx, non-targeted neuromodulation of the nerve causes off-target effects, which limits the applicability of such devices. Human vagal nerve stimulators for epilepsy, for example, cause cough, voice alteration, nausea, and pain [16], which are evoked by activation of pulmonary afferents, recurrent laryngeal efferents, and pain nociception fibres. A study into the effectiveness of treatment for chronic heart failure concluded that effective treatment cannot be delivered without side effects such as dysphonia, cough and stimulation-related pain. These necessitated the reduction of the stimulation intensity which in turn reduced or eliminated the therapeutic effect [9].

A potential way of eliminating side effects would be to deliver the current in a spatially selective way, where only the specific part of the nerve containing the required nerve fibres is activated. This can be achieved through either current steering with an external cuff device [17]–[19] or penetrating electrode arrays [20], [21]. The latter is currently far away from clinical applications because of concerns regarding their chronic use; the presence of rigid foreign material in nerves causes tissue damage and inflammatory response [22], [23]. The application of flexible extra-neural multi-electrode devices on vagus nerves to aid selectivity of recruitment is a viable option, but it is still at an early stage. The most recent advances demonstrated blood pressure control without side effects in rats [24], and selective neuromodulation of cardiac, pulmonary, urinary, and endocrine function in dogs [25]. There are also some preliminary results on selective alteration of heart function in pigs [26] and humans [27]. However, those studies used arbitrary electrode geometries and count; electrode arrays were reportedly difficult to fabricate, and the latter studies were performed in two dogs and four humans, respectively, with inconsistent results on efficacy and inconsistent physiological alterations reported.

Hence, there is a clear need for the investigation and optimisation of such extra-neural multi-electrode cuff arrays, which can be surgically implanted around peripheral nerves, such as the vagus nerve, and can potentially be used for multiple disease treatments by altering the spatial selectivity of the current to the fascicle of interest.

In terms of the applied stimulation waveform, there is a multitude of pulse waves which can be employed to increase the efficiency of the neuronal stimulation, ranging from unidirectional square to decaying or raising exponential [22], [23]. Most human-approved devices, however, use a biphasic charge-balanced waveform due to the safety reasons [30] and the fact that an implantable pulse generator with such high-resolution waveforms is highly energy inefficient and difficult.

Total multi-parametric optimisation of the device is very difficult in this case due to the number of variables involved in the optimisation process. Several studies have demonstrated optimisation of a certain subset of parameters while fixing other parameters for various reasons. Specifically, in [31] a method was developed which evaluated optimal multi-polar stimulation patterns and applied waveforms, while the geometry of the device itself was fixed. The authors used generic human-sized vagus nerves, and demonstrated the effectiveness of optimisation *in vivo* and chronically. In [32] the number of contacts within the cuff array was studied, and the relative position of the contacts was addressed in [33] However, while these studies showed promising results, the detailed geometrical optimisation of the multi-electrode cuff itself has never been performed before.

The purpose of the presented work was to optimise the geometrical parameters of the array in a generic nerve model, while waveform and pulse duration was fixed, and test the efficiency of the optimised array *in vivo* in human-sized vagus nerves (sheep).

Particular questions were: a) What are the optimal geometry, configuration, and the number of electrodes in the cuff? b) Does the optimal device work in *in vivo* acute animal experiments?

## 2. Methods

### 2.1 Experimental Design

As mentioned above, we employed a biphasic symmetric square pulse as this is a reliable standard used in implantable stimulators [30]. The optimisation of the array geometry was performed in simulations where a 3D FEM model of a human-sized nerve was used to compute static current and voltage distributions. It was further coupled through an activating function to 1D time-dependent FEM models of the myelinated axons, running inside and parallel to the nerve, which was used to simulate action potential (AP) initiation and propagation at any given cross-sectional location [34]. During optimisation, we targeted specific axons inside the nerve in two locations. These fibres represented the stimulation goal, while all other fibres within the nerve represented the unwanted side effects: their stimulation was not desired. We explored a wide range of geometric parameters and in each case, we have iteratively computed the threshold current required for those given axons to fire an action potential. Then we looked at the entire cross-section of the nerve and identified for every location whether the fibre placed in that location would fire an action potential. Then we have computed an objective cost function (OCF) resampling combination of two unwanted 'side effects' of targeted stimulation: 1) total cross-sectional area of the nerve where axons have been activated in addition to the target axon, and 2) threshold current density directly under the electrodes. The combined influence of both effects was computed as a normalised sum and minimised.

The rationale behind the selection of the OCF is the following: We would like the targeted fibre within the nerve to be stimulated, at the same time we would like no other fibres within the nerve to be active, and we would like to do it with the minimal possible current. As the electrodes become smaller, the expected selectivity increases. At the same time, the current density underneath the electrodes increases which is associated with the increased tissue damage and lower selectivity margin: the gradient of the current flow becomes higher and the spread of current increases towards the electrode edges. Higher current also requires more energy of the IPG to be delivered to the tissue which is also unwanted. Proposed OCF naturally finds the balance between minimising the amount of the cross-section of the nerve being stimulated and minimising the amount of current required for the targeted fibre stimulation.

First, we assumed symmetric configuration of one pair of electrodes along the nerve and optimised their dimensions and the distance between them. Then we showed that an optimal symmetric configuration gave the best results among any (non-symmetric and symmetric) configuration, and, in particular, demonstrated its superiority against a multi-electrode configuration. During the last stage of this process, we calculated the optimum space between the symmetric pairs along the circumference of the nerve and formulated the final specifications for the optimum cuff electrode array.

The suggested optimal design was adjusted given real-life manufacturing and impedance considerations and then 12 devices were manufactured and implanted around the cervical vagus nerve in 12 healthy anaesthetised sheep. Electrical stimulation was performed on each of the 14 electrode pairs in the array in turn; thus, activating 14 spatial sections around the nerve while measuring respiratory breath rate (RBR) through end-tidal $CO_2$ ($EtCO_2$), and heart rate (HR) through an electrocardiogram (ECG) and arterial blood pressure (ABP). Sites with significant changes in respiratory or cardiac rate were identified. The stimulation current amplitude was then adjusted if necessary with the aim of selective activation of HR without effect on the RBR and vice versa.

*2.2 Model-based optimisation*

A 3D cylindrical model of a human-sized vagus nerve was produced in the COMSOL simulation software (COMSOL Ltd, UK). The model was made without fascicular organisation as the this varies from species to species and will be unknown *a priori* for a particular subject. Previous studies have also shown a negligible effect of fasciculation and applicability of a generic model for accurate predictions in the experiments [31]. The model was 2.8 mm in diameter and had two compartments: intraneural space (homogeneous conductivity of 0.3 S/m) and 100 μm-thick epineurium (0.083 S/m, [35]) surrounding the latter (Figure 1).

Discretisation was performed according to mesh convergence criteria resulting in the optimal mesh to be 5M regular tetrahedral elements with the refined area near the application of the electrodes. This mesh converged for the smallest electrode size with the tolerance of $<10^{-6}$ in all nodes with respect to the voltage estimation. The electrodes were modelled using normal current density boundary conditions with surface contact impedance (so-called complete electrode model [36]), which accounts for non-homogeneities of the current distribution due to contact impedance. The impedance of the electrodes was set to 500 Ohm·mm$^2$ to match experimentally evaluated impedances for actual electrode arrays [37]. The external boundary conditions were supplied as a constant normal current density on the electrodes. The problem was solved in quasi-static formulation (Figure 2, block 1).

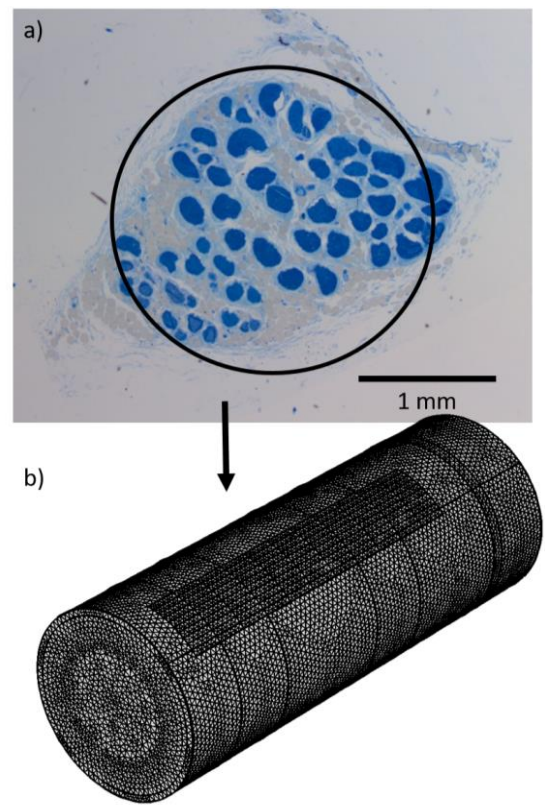

Figure 1. a) A representative semi-thin histological section of a sheep vagus nerve, and b) the generic model of the human-sized vagus nerve used for the optimisation process.

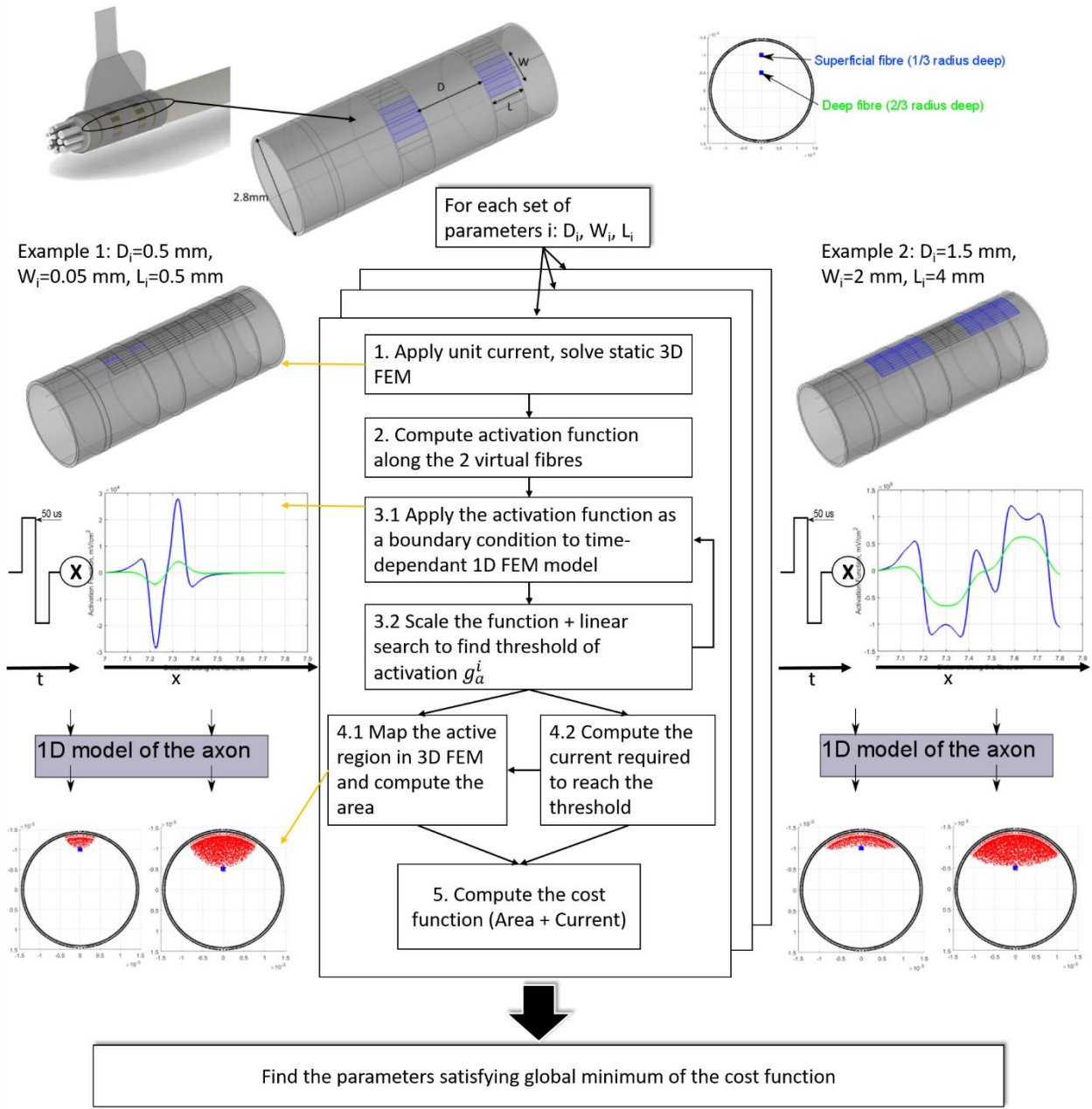

Figure 2. Experimental design and overview of the model-based optimisation methods for step 1 (symmetric longitudinal configuration). A generic homogeneous 3D model with two compartments (epineurium and intra-neural space) of the nerve was used to study the effects of electrode geometry and optimise geometric parameters according to the flowchart. The quasi-static fields were computed using this model (1), from which the representative shapes of the activating function were extracted (2) and applied as a boundary condition to the time-dependant model of the myelinated axon (3). Using the iterative linear search, the threshold of activation was detected, the threshold current was computed, and active areas were mapped back to the 3D model (4). After this, the OCF was computed for each geometrical configuration and studied. Minimal OCF indicated optimal geometrical parameters. Two exemplar parameter sets are presented for illustrative purposes.

Two radially located "virtual fibres" were placed beneath the electrodes, one at a third and another at two thirds of the radius towards the nerve centre, to serve as a target for neuronal stimulation. The simulations were performed for each set of parameters $p_i$ (Table 1), supplying unit current between the electrodes and evaluating the voltage distribution in the volume $V$. Then the activating functions were computed along the virtual fibre locations [38]:

$$g(x) = \frac{\partial^2 V(x)|_{y=y_0, z=z_0}}{\partial x^2} \quad (1)$$

where $x$ was the longitudinal direction parallel to the axon, and $(y_0, z_0)$ were the cross-sectional coordinates of the fibre.

The activating function was multiplied by a temporal biphasic square pulse (positive first) with 50 μs pulse width (Figure 2, block 2,3). The resulting spatiotemporal function was applied as a boundary condition to the time-dependent model of a myelinated axon.

An accurate 1D FEM double-cable model of a mammalian myelinated sensory fibre was used as a fibre model [39]. It contained ten ion channels taken from an experimentally validated space-clamped model [40]: four at the node and six at the internode. A realistic morphology of the fibre [41] was implemented similarly to the one used in [42]. The fibre full diameter was 15 μm and the model was implemented within the same simulation environment in COMSOL (COMSOL Ltd, UK). The conduction velocity of the fibre and the shape of membrane potential at the nodes and internodes in resting state and during excitation matched an experimentally validated space-clamped model [40]. The complete description and validation of the model is presented in [39] and the source code is freely available in [43].

The amplitude of the activating function was varied in order to find the threshold of activation. It was determined for each set of parameters using an iterative linear search.

After the threshold $g_a$ of activation for the activating function was established, the threshold current needed for activating the fibre was recalculated for each parameter $p_i$. Then the total activated area in the cross-section (above the activation threshold) $A(\|g(x)\| > g_a)$, and maximum current density directly beneath the electrodes $J_i$ were calculated. The normalised sum was computed as an objective cost function (OCF) and minimised over the parameters (note that the maximum active area equals to the half of cross-sectional area of the nerve $\frac{1}{2}\pi R^2$, where $R = 1.4$ is the nerve radius):

$$F_i = \frac{J_i}{\max_i \|J_i\|} + \frac{A(\|g_i(x)\| > g_a^i)}{\frac{1}{2}\pi R^2} \quad (2)$$

$$p_{opt} = \frac{\operatorname{argmin}(F_i)}{P_i} \quad (3)$$

OCF was computed for both superficial and deep fibre, and combined OCF was computed as an average between the two. The combined OCF was used for global optimisation throughout the full parameter space. Some results with individual OCF for separate fibres were projected over the single parameter and were presented to analyse the trends and aid deeper understanding.

**The optimisation was performed in three steps:**
**Array optimisation step 1:** Optimisation of symmetrically located pair of electrodes.

Before considering more complex geometrical arrangements, a symmetrical longitudinal bipolar configuration was optimised. Here, a single pair of identical electrodes was placed longitudinally along the nerve on the same radial "o'clock" position. Electrode width, length, and distance between the electrodes were varied using the range of parameters in Table 1. Once all permutations of parameters were computed (total 10 000) the best combination according to the criteria (3) was selected.

**Array optimisation step 2:** Proof that the symmetric configuration is optimal among all other geometric arrangements of two electrodes (Figure 3).

Here, given the symmetry of the problem, one of the electrodes was fixed with its optimal parameters, while the other was changed: the width and length of the electrode, as well as its radial and longitudinal location, were increased from half of the optimum value to double. The OCF (2) was computed along with the change of each individual parameter. The OCF increased per each permutation indicating that the optimal parameters from step 1 satisfying criteria of at least local minima over more general space of geometric configurations.

To illustrate this point further, and to evaluate the effects of using multiple contacts with the regular grid arrays, the cost function was computed for several examples previously used in the literature: 1) 2-electrode cross-sectional adjacent, where electrodes were located in the same longitudinal position but different "o'clock" locations (Figure 3, bottom left), and 2) multi-electrode. We explored 2 configurations of the multi-electrode current spread. In both, one of the electrodes was fixed and the other was replaced by an array of 3, 6, 9, or 13 electrodes. It was either within the same electrode "ring" (Figure 3, bottom centre) or within the "ring" shifted by the optimal distance. (Figure 3, bottom right). The geometry of each electrode in the multi-electrode part was equal to the optimal geometry and the electrodes were spaced with the optimal configuration grid spacing form step 1.

**Array optimisation step 3:** Computation of radial spacing between the pairs in the array.

Once the optimal parameters were computed and symmetric configuration was proven optimal, the spacing between the pairs, or their respective "o'clock" positions, was computed by half of the length of the maximum spread of the activated area $A(\|g_i(x)\| > g_a^i)$. In other words, for the given electrode pair, the neighbouring pairs were placed so that they were on the edge of the activation area.

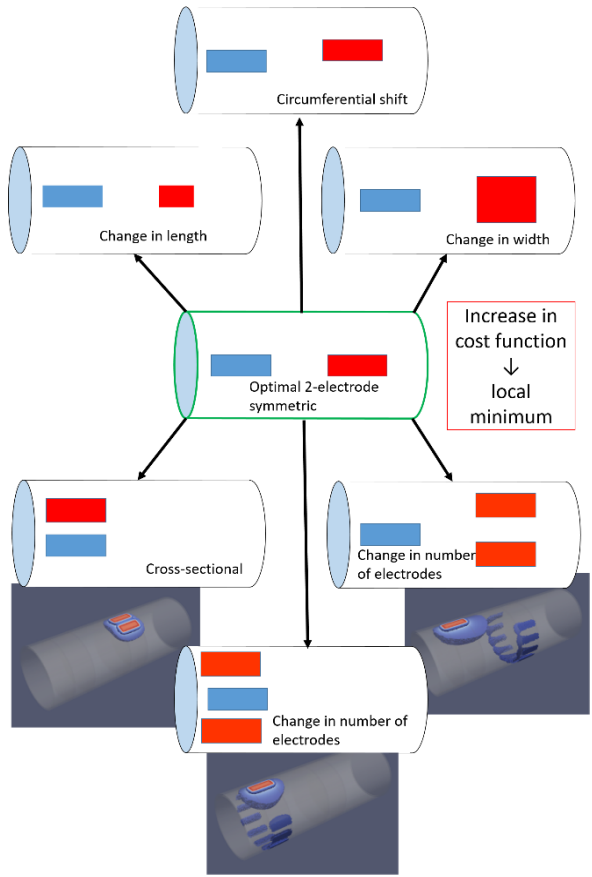

Figure 3. Schematic diagram of step 2 of the array optimisation. The optimisation of the symmetric longitudinal pair (Figure 2) was taken as a starting point (middle) and then each of the geometric parameters (length, width, radial location, and distance) was perturbed around this starting location for one electrode. The OCF was increasing in response to change in any parameter in any direction. Hence, the starting configuration satisfied at least local minimum. Three alternative common configurations were tested (see 2.2) with a similar increase in OCF.

Table 1. Ranges of electrode parameters used for the optimisation

| Parameter | Width (W) | Length (L) | Distance between electrodes (D) |
|---|---|---|---|
| Range | 0.05 – 2.0 mm | 0.5 – 4.5 mm | 0.3 – 4.5 mm |

Selected optimal parameters were then slightly adjusted given the practicality of the manufacturing and *in vivo* experimental requirements (mainly laser cutting accuracy of 0.03 mm, PDMS stiffness, and real-life contact impedance after PEDOT: pTS coating) and optimal designs were produced as described in the next section.

## 2.3 Electrode array fabrication

Optimal electrodes were fabricated using a laser cutting stainless steel-on-silicone rubber approach. Connecting tracks were 150 μm wide with an inter-track spacing of 35 μm. The total thickness of the array was 220 to 250 μm. After fabrication, electrodes were coated with PEDOT:pTS to reduce contact impedance below 1 kOhm at 1kHz for each electrode [37].

The following protocol was used (Figure 4): 1) A 10 μm release layer (Poly(4-Styrenesulfonic Acid), Sigma Aldrich, UK) was spin-coated on the 70x40 mm glass slides (Fisher Scientific, UK) followed by approximately 100 μm layer of the silicone rubber (MED4-4220, Polymer Systems Technology Ltd, US). 2) The silicone was cured for 5 mins at 100°C in the oven. 3) Stainless steel foil (0.0125 mm, Advent Research Materials, UK) was glued on top of the rubber. 4) The electrode profile was cut using a 20 W precision laser cutter (355 nm, IDG lab, UCL, UK). 5) Excess foil was peeled off, and the second 100 μm layer of the silicone rubber spin-coated on top. 6) The electrode openings and connector pads were laser-cut and peeled away. 7) Finally, the array was glued on the inside of the 3 mm ID silicone rubber tube (Sigma-Aldrich, UK) with thread mounted to assist implantation and the connector side was attached to a custom PCB using three 20-way 1.5 mm connectors (Farnell, UK). The complete fabrication process, together with the evaluation of the electrode performance, can be found in [37].

## 2.4 In vivo experiments

*2.4.1. Anaesthesia and Maintenance.* Twelve sheep were anaesthetised by induction using ketamine (5 mg/kg) and midazolam (0.5 mg/kg) injected IV into the jugular vein. A tracheal tube was then inserted into the trachea for the primary purpose of establishing and maintaining a patent airway and to maintain general anaesthesia using sevoflurane carried in an oxygen/air mixture. After induction of general anaesthesia, the animal was positioned in dorsal recumbency. Indwelling catheters were percutaneously placed in both the external jugular veins and one in the femoral artery (for blood pressure and blood gas monitoring) using ultrasonographic guidance. An intra-oesophageal tube was inserted to drain refluxes from the rumen. The animal was instrumented with ECG leads and a pulse oximeter. A spirometer was connected to the tracheal tube. An intranasal probe was used to control the temperature.

*2.4.2. Surgery.* The palpebral and corneal reflexes, medioventral eye ball position, and jaw tone were used to monitor anaesthetic depth prior to surgery initiation. Nystagmus as well as lacrimation were monitored as possible signs of a light plane of anaesthesia. The animal was mechanically ventilated using a pressure control mode for the duration of the surgery and implantation, after which the animal was kept on spontaneous ventilation during the stimulation sessions. Animals were placed onto mechanical ventilation between rounds of stimulations if required to restore normal levels of $CO_2$ (between 35-45 mmHg). Body temperature was maintained using a hot air warming system if necessary. Ringer lactate fluid therapy at a rate of 5 ml/kg/h was administered intravenously throughout the procedure. Glucose was administered at the discretion of the anaesthetist when blood glucose was lower than 4 mmol/L.

*2.4.3. Monitoring and recordings.* Arterial blood gases were analysed using a blood gas machine (i-stat, Abbott, USA) throughout the experiment to monitor pH, Glucose, $PaO_2$, $PaCO_2$, Bicarbonate, $SaO_2$, $Na^+$, $Cl^-$, $Ca^{2+}$ and $K^+$ levels. All physiological parameters (including HR, systemic arterial blood pressure (sABP), central venous pressure) as well as the level of delivered sevoflurane were recorded. Depth of anaesthesia throughout the experiment was assessed by inspecting physiological parameters as well as using a bispectral index monitoring system (BisTM, Medtronic, USA). Levels of sevoflurane were adjusted accordingly by the anaesthetist. In some cases, boluses of propofol or fentanyl were used, if required, and noted on the records. Physiological data were outputted from the anaesthesia monitor (Aestiva 5, General Electric, Finland) into a 16-channel Powerlab system. Raw ECG, systemic arterial blood pressure (sABP), end-tidal $CO_2$, % of expired Sevoflurane, and central venous pressure were digitalised by a Powerlab acquisition system and sampled at 2000 Hz with LabChart 8.0 software (AD Instruments, UK). Calculated channels (for live measurement of HR, systolic ABP, diastolic ABP, mean ABP, respiratory rate) were added in the Labchart as needed. A trigger channel was also used to link the stimulator to the physiological data.

*2.4.4. Electrode implantation.* After induction of anaesthesia and placement in dorsal recumbency, the ventral neck region was clipped and aseptically prepared using chlorhexidine-based solutions, prior to the placement of sterile drapes, leaving only a small window open for accessing the right cervical vagus nerve. Using aseptic technique, a 20 cm longitudinal skin incision was made using monopolar electrocautery centred immediately to the right of the trachea from the larynx caudad. The incision was continued through the subcutaneous tissue and the sternohyoideus musculature using a sharp/blunt technique until encountering the carotid sheath and right vagus nerve. Two segments measuring 3-5 cm each of the right vagus nerve (RVN) were circumferentially isolated by blunt dissection. Artery forceps were then inserted under the nerve from lateral to medial, grasping one pair of suture glued to the custom made cuff electrode array introduced into the surgical field. The cuff was placed around the nerve by pulling on the opposite pair of sutures to open the cuff and carefully sliding the vagus inside it. Electrical ground and heart electrodes were inserted into the surgical field. The impedances of the electrodes were tested to make sure they were less than 1 kOhm at 1 kHz. The surgical field was then rinsed with sterile saline and the skin temporarily closed using towel clamps.

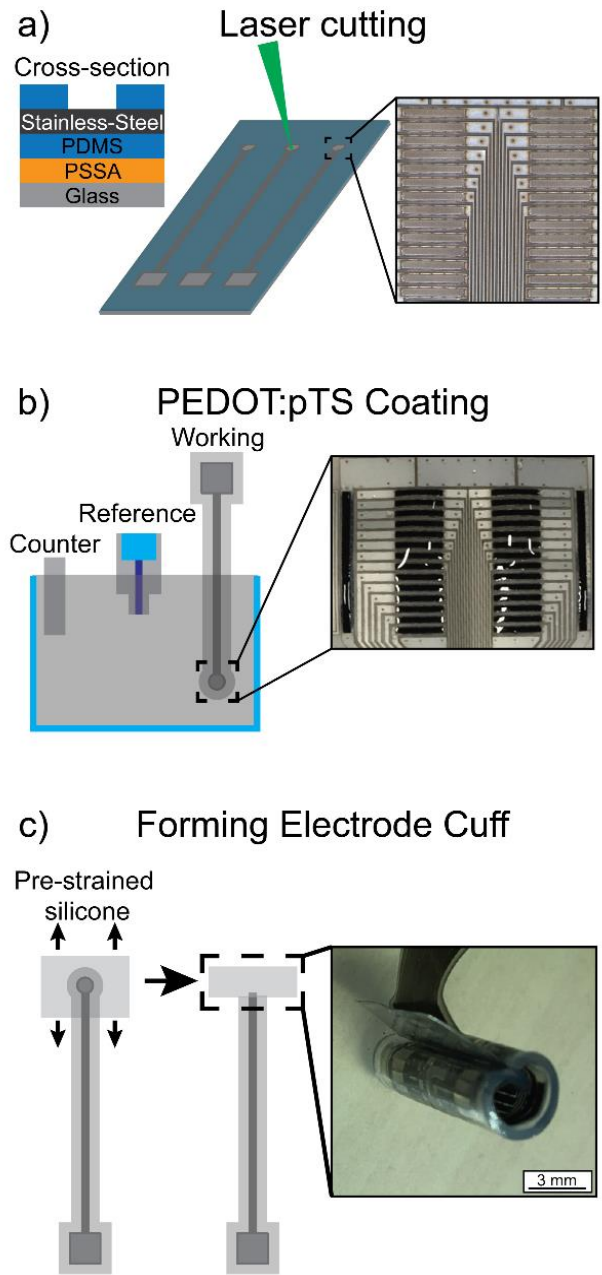

Figure 4. Electrode fabrication procedure. a) The stainless steel foil is attached to a PDMS substrate, laser-cut, coated with the second layer of PDMS and patterned following electrode sites opening. b) Electrode sites are coated with PEDOT:pTS. c) The cuff is formed into the tubular configuration by either using existing PDMS tube or glued to a sandwiched pre-stressed PDMS sheets.

*2.4.5. Electric stimulation.* The optimised arrays were placed around the vagus nerve during terminal experiments in 12 healthy sheep (Figure 5). A protocol of stimulation with the multi-electrode array was generated using MATLAB. Stimulation was delivered to the RVN using one pair of longitudinally paired electrodes at a time. Each stimulation consisted of a period of 30 s of stimulation and 30 s of recovery. Matched longitudinal pairs in the same radial position on both rings were sequentially selected for current application. All the stimulations were performed while the sheep was at steady-state and the same depth of anaesthesia. Physiological endpoints were measured (EtCO$_2$, ECG, ABP) from which HR and RBR were calculated. During the experiment, the current amplitude was adjusted in 100 µA steps so that: a) significant (more than 5%) physiological effects of stimulation for both RBR reduction and HR reduction were present in at least one out of the 14 pairs, and b) the effects do not intersect, i.e. the stimulation of any 'RBR'-sensitive pair does not cause HR changes and vice versa. The selected amplitude of stimulating current was recorded and used throughout for reproducibility. The observed responses for each stimulation pair were displayed, evaluated and compared with the baseline (30 sec prior to stimulation). Stimulation was delivered with a square biphasic (anodic first on the proximal contact) constant current temporal waveform using a balanced current source (Keithley, UK model no 6221, with a pulse width of 100 µs; 50 µs for each phase and no interphase delay). The frequency of pulses was 20 Hz.

In 4 sheep the stimulation with current amplitudes 100-900 µA with steps of 100 µA were performed in all possible stimulation pairs. The sensitivity was computed as a percentage of stimulation pairs which elicit the significant physiological response (change in RBR) and analysed with respect to the current amplitude. In those 4 sheep, the whole vagus nerve was also stimulated with the amplitude of 5 mA using the guard electrodes to compare the selective and non-selective physiological responses.

*2.4.5. Statistical analysis.* The two-sided t-test was used to access the statistical significance of the results within the subject. The repeated measures one-way ANOVA was used to access the significance of the results across the animals. All results are presented as Mean ± 1 SD if not stated otherwise.

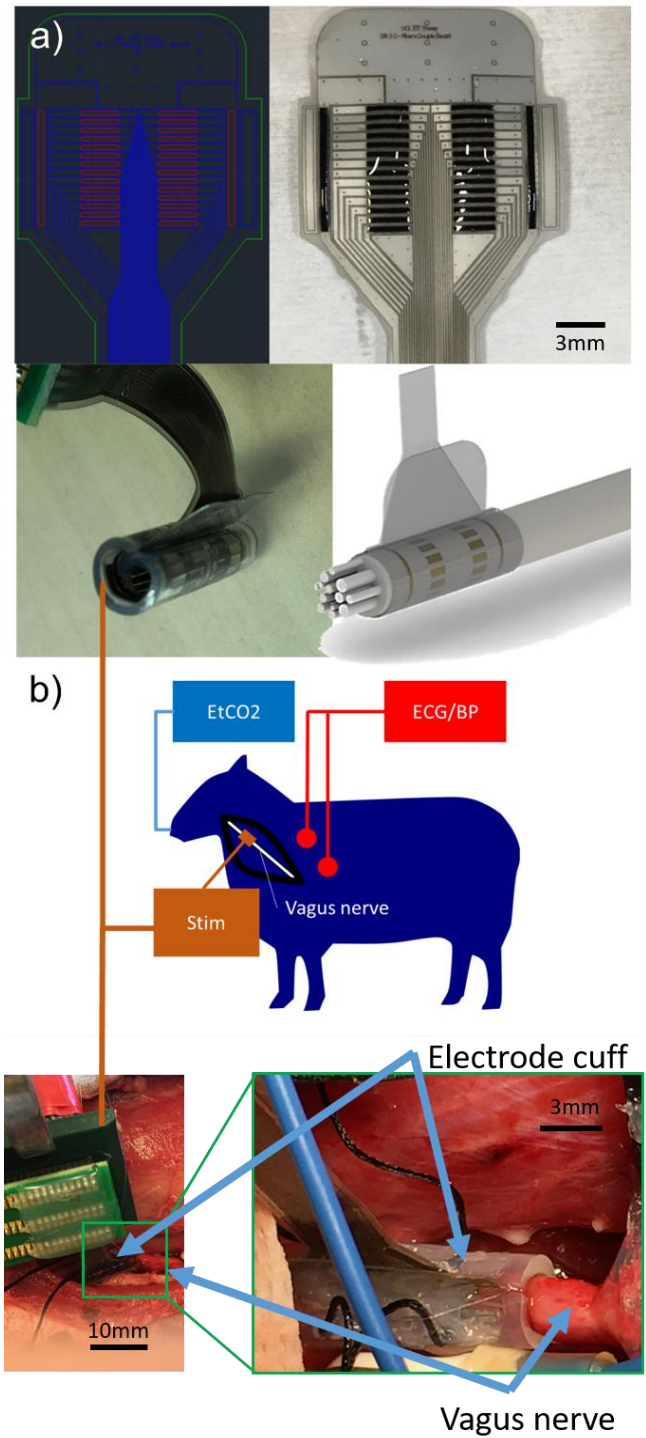

Figure 5. The experimental setup for *in vivo* evaluation. a) An optimal electrode array was manufactured and b) implanted on the vagus nerve of an anaesthetised sheep. Physiological parameters were measured whilst probing the circumferential electrode locations with 30 s stimulation followed by 30 s interstimulus time. The bottom figure shows the placement of the cuff electrode array on the vagus nerve.

## 3. Results

### 3.1 Optimisation in simulations

*3.1.1. Optimisation step 1.* The current density beneath the electrodes exponentially decreased with electrode area (Figure 6.d). Despite this, however, the total injected current required for fibre activation did not change exponentially with electrode area (Figure 6.e). It was 350±20 µA for the deep, and 160±10 µA for the superficial fibres with the smallest electrodes (0.05 mm$^2$), compared to 420±10 µA and 200±10 µA, respectively, for the largest (2 mm$^2$). The influence of the geometry was apparent in the longitudinal shape of activating functions (Figure 6.c). This subsequently altered the threshold of activation, with the most influential parameter being the distance between the electrodes (Figure 7.a): for the shortest electrodes (0.5 mm) located close together (0.5 mm) it was 9.9±0.1*10$^7$ mV/cm$^2$ which was halved (5±1*10$^7$ mV/cm$^2$) when the same electrodes were located 4.5 mm apart. Sensitivity analysis of the OCF around the globally optimal configuration showed that the electrode length had the least influence on the OCF (*Table 2*), although the general trend was that the deeper fibre was activated more efficiently with longer electrodes (OCF decreased by 6% per 25% increase), while shorter electrodes worked better for superficial fibre (OCF decreased by 5% per 25% decrease).

The increase of inter-electrode distance led to drop in the threshold of fibre activation (Figure 7.a) because of the increase of the total transfer impedance between the electrode, with the subsequent increase of the voltage difference and therefore the amplitude of activating function for the same injected current. The additional aspect was the increase of redistribution of the current density towards the electrode edges which also increased the amplitude of activating function.

The most significant parameters for optimisation were the electrode width and the distance between the electrodes (Figure 7.b, *Table 2*), which changed the OCF value significantly across the range of configurations. Electrodes wider than 0.8 mm produced a significant artificial drop in OCF (Figure 7.b, drop by 0.1 ± 0.05 for all configurations): reduction in current density on the electrodes was no longer compensated by an increase of the affected area, as at this point it was equal to half of the nerve and could not increase further. These configurations were therefore discarded.

The absolute minimum of the combined OCF was with 0.4 mm width, 0.5 mm length, and 3 mm distance between the electrodes. However, after evaluating the manufacturing constraints (accuracy of the laser cut and overall stability of the device) and considering contact impedance value as an additional parameter (longer electrodes produced lower impedance, but did not significantly alter OCF (Figure 7.b), it was decided to select another configuration, which was in the top 1% but had much longer electrode length (Table 2).

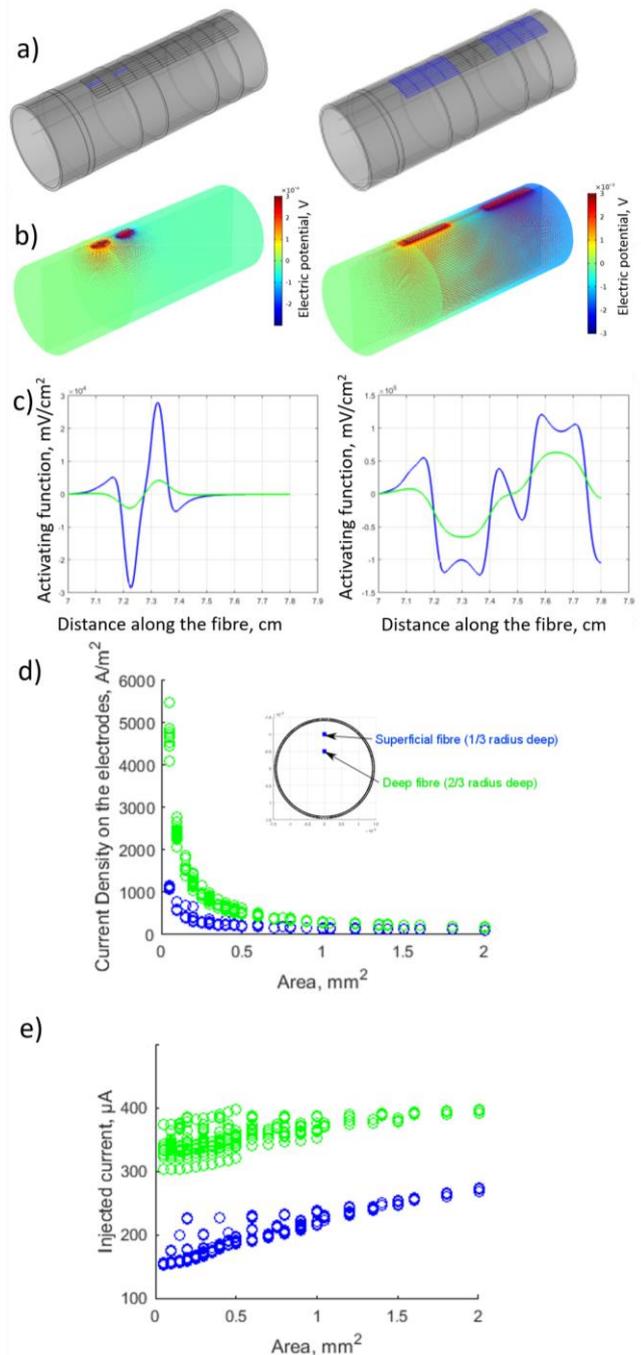

Figure 6. a) Two examples of simulated bipolar symmetric electrode pairs. b) The potential (colour) and current density (red vectors) distribution within the simulated volumes of the nerve section. c) Activating functions for the above cases, taken along the deep (green) and superficial (blue) fibre locations. d) Threshold current density beneath the electrodes for activating deep (green) and superficial (blue) fibre VS total electrode area. e) Threshold current for activating deep (green) and superficial (blue) VS total electrode area.

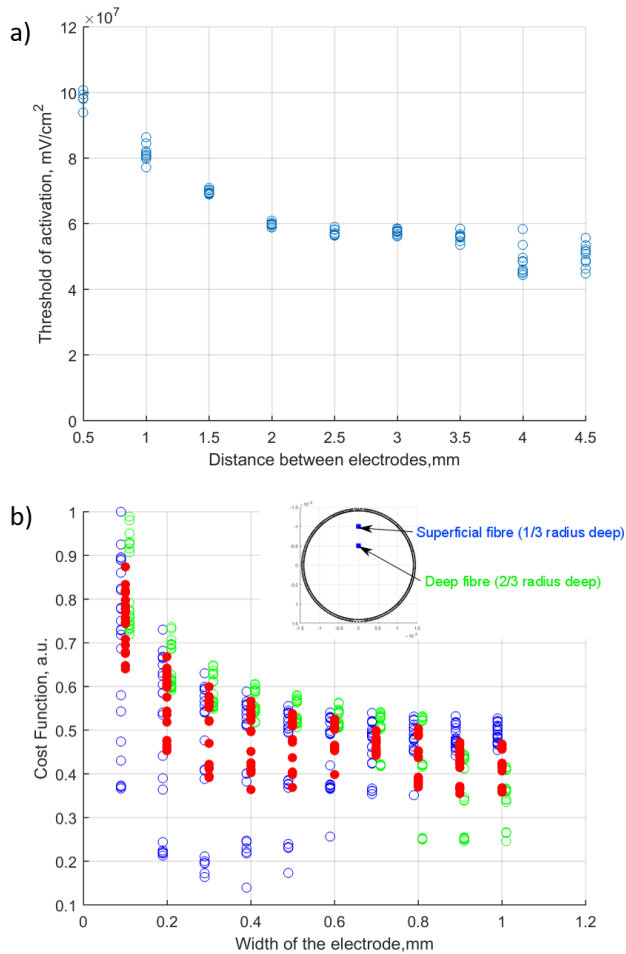

Figure 7. Summary of simulation results with bipolar longitudinal electrode configuration. a) The threshold of activation variation with respect to the distance between the electrodes for the shortest (0.5 mm) electrodes. b) OCF vs. electrode width for the deep (green), and the superficial (blue) fibres, and combined (red).

Table 2. Parameters sensitivity analysis around the optimal configuration

|  | Optimal −25% OCF | Optimal OCF | Optimal +25% OCF | Average Sensitivity |
|---|---|---|---|---|
| Distance | 0.39 | 0.36 | 0.40 | 10% |
| Width | 0.39 | 0.36 | 0.38 | 7% |
| Length | 0.37 | 0.36 | 0.36 | 1% |

*3.1.2. Optimisation step 2.* The optimal bipolar longitudinal configuration evaluation was followed by the proof of the hypothesis that this configuration is optimal among any geometric electrode geometries. Since the problem is symmetric, one of the electrodes was fixed in its optimal state, whilst the second was changed with the simultaneous objective function calculation. The OCF was 5% higher for 10% width variation, 1% higher for 10% length variation, and 1% higher for 10% radial position variation.

When considering the adjacent configuration, the OCF for any combination of the electrodes was more than 50% higher. For multi-electrode configurations, the OCF increased by 30%, 40%, 44%, and 45% for the cross-sectional multielectrode configuration, and 33%, 41%, 44%, and 45% for the longitudinal multielectrode configuration, when the current was spread over the 3, 6, 9, and 13 electrodes respectively, thus offering no advantages to the 2-electrode configuration.

*3.1.3. Optimisation step 3.* For the full multi-electrode array layout, the inter-electrode circumferential distance – or the distance between neighbouring pairs for the optimal geometry – was 0.35 mm. The electrode width was then slightly adjusted so that the complete circumference of the nerve was covered leading to 14 electrode pairs around the 2.8 mm-diameter nerve (Figure 5.a).

Electrodes were successfully manufactured; contact impedance values in saline were 297 ± 1.04 Ω at 1 kHz (n = 28 individual electrodes).

Table 3. Optimal parameters of the simulated array, and actual manufactured device after consideration of production constrains

| Parameter | Optimal design | Manufactured near-optimal device |
|---|---|---|
| Width | 0.4 mm | 0.35 mm |
| Length | 0.5 mm | 3 mm |
| Distance between the electrodes | 3 mm | 3 mm |
| Number of electrodes | 14 | 14 |
| Percentile of OCF across all other combinations | best (OCF = 0.36) | Top 1% (OCF = 0.37) |

### 3.2 In vivo evaluation

Reproducible changes in both RBR and HR in response to stimulation occurred in all 12 anaesthetised sheep (P < 0.01, Figure 9). Sequential stimulation with the increasing current was applied until a clear bradypnea response was observed in at least one pair of electrodes. The stimulation current was then adjusted to maintain the significant response (reduction in RBR) induced during stimulation through at least one electrode pair while making sure RR and HR responses do not intersect, i.e. the RBR-sensitive stimulation pair does not cause HR changes and vice versa. This optimal current amplitude was 500 ± 50 μA (N = 24 in 12 sheep).

In this condition, a separate area evoking a cardiac response (reduction in HR) to stimulation could be identified in 9 out of the 12 animals. In three out of the 12 sheep, the HR did not change significantly during stimulation through any of the pairs tested even after a further increase in current amplitude which caused RBR response in all 14 pairs.

In 4 sheep the current sensitivity analysis was performed with the 100 μA step size. The optimal current which elicits the physiological response in at least 1 pair (7%) was found to be 500 ± 60 μA. Increasing the current to 900 uA elicited the significant RBR and HR response in more than 6 out of 14 of the pairs for all 4 sheep in RBR case, and 3 sheep in HR case (Figure 8, for HR case in 1 sheep no HR response was observed for all stimulation amplitudes).

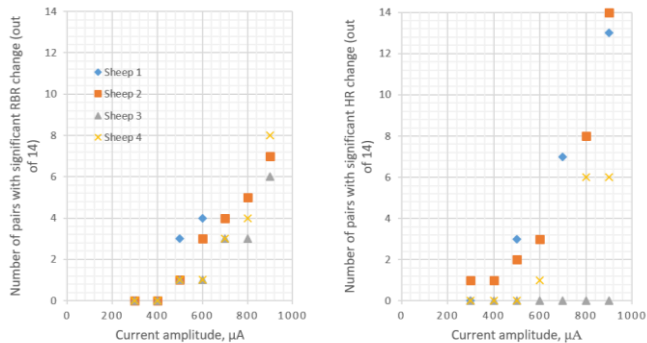

Figure 8. Selective current sensitivity analysis. The number of pairs which evoke the significant physiological response (RBR-left, HR-right) VS the current amplitude (N=4 in 4 sheep). The optimal current amplitude was 500±60 μA across all animals (N=24 in 12 sheep).

Across all animals, the RBR was reduced by 85 ± 5% with no change in HR (P > 0.5) when stimulating through the pair of electrodes evoking an optimal bradypnea response with optimal current amplitude in all sheep. The HR was reduced by 20 ± 7% with no change in RBR (P > 0.5, repeated measures ANOVA) when stimulating through the pair of electrodes evoking optimal cardiac response (N = 24 in 12 sheep).

When comparing the selective VS non-selective stimulation, in 4 control sheep there was no significant difference with respect to the size of the RBR and HR change between stimulation of the whole nerve and selective stimulation (P>0.1, repeated-measures ANOVA, n=8 in 4 sheep). RBR changed by 90 ± 10% for all 4 sheep, and HR changed by 20 ± 10 % for 3 out of 4 sheep. In 1 sheep both whole-nerve stimulation and selective stimulation did not cause any significant change in HR irrespective of the current amplitude.

The data from all animals were grouped together by rotational alignment by the best respiratory-responsive site. The %RBR and %HR change, with respect to the baseline, was calculated for each dataset, normalised across all animals, and plotted on the diagram. This revealed significant spatial localisation of respiratory- and cardiac- responsive sites with respect to each other (P < 0.01, repeated-measures ANOVA, Figure 10) with an angular position of 105 ± 5 degrees.

## Discussion

### 4.1 Summary of results.

Optimisation of the electrode geometry in simulations allowed objective determination of geometrical parameters for selective neuromodulation of myelinated fibres in the vagus nerve. The best geometric configuration, given the biphasic square pulse application, was the symmetric longitudinal, same "o'clock" pair. The optimal electrode array had 14 longitudinal electrode pairs, located 3 mm apart. The optimal width was 0.4 mm, with 0.35 mm inter-electrode circumferential distance, and a length of 0.5 mm. The latter was found to not significantly alter the objective function and was replaced by 3 mm (with minimal penalty) due to contact impedance and manufacturability considerations.

Further analysis showed that using non-symmetric electrode geometries, as well as multi-electrode geometries, does not improve OCF when they are used with a bipolar current source.

The arrays were manufactured and tested in healthy anaesthetised animals, where the stimulation current was adjusted to ensure spatial selectivity of the stimulation. The resulting current amplitude of 500 ± 50 μA for this array was not too far from that predicted by simulations, which was between 200 and 400 μA depending on the targeted fibre location. The detected mismatch is likely due to the thin layer of saline between the electrode and the nerve, which was not modelled. This might also explain the variance between animals (the range was between 200 and 800 μA) as the nerve thicknesses varied from animal to animal. This, however, should not significantly modify the optimisation results as the generic model scales the electric field distribution linearly and, although the total current is different, the OCF converges to a minimum at the same geometrical configuration. Results also demonstrate the potential of the generic model to be a good compromise when unknown *a priori* nerve geometry is under investigation.

Reliable and reproducible selectivity achieved for the neuromodulation of respiration-specific and cardiac-specific function (100% spatial separation between them, Figure 10) has never been reported before. At the same time, the current that was optimal for the best selectivity was not far off that predicted by the model, even though the model was not tailored specifically for each animal. Finally, the current sensitivity analysis shows the increased margin of control which is predicted by the model (Figure 8). Overall, the experimental study demonstrates that the array works and can be used for selective stimulation, at the same time achieving the degree of selectivity in the vagus nerve which was never demonstrated before. This suggests that the proposed generic model-based approach could be applied to optimise the geometry for cuffs of different diameters.

These experimental results also suggest the spatial organisation of effect-specific fibres in the cervical vagus nerve even in large animals. The described effects could be caused by organotopic organisation of cervical vagus nerve, or by the spatial organisation by efferent/afferent fibre types as demonstrated in [44], given the fact that the HR change is primarily an efferent effect, and RBR is the reflex caused by the afferent fibres stimulation [45]. In the present study, the type of the symmetric biphasic pulse used did not allow us to distinguish efferent/afferent stimulation as the action potential was generated in both directions [46], [47], so further studies are required to fully understand the nature of this spatial organisation.

From a clinical perspective, all of the above also demonstrated that cardiac-specific and/or respiration-specific neuromodulation is possible without inadvertent stimulation of the other function. A potential application of this approach would be a refined vagus nerve stimulation method which targets specific areas of the nerve while avoiding the stimulation of unwanted response transmitted to other organs.

In addition, even if the particular therapeutic effect of vagus nerve stimulation is difficult or impossible to quantify acutely, the specific cardio-respiratory nerve sections could be identified by trial-and-error and the current could be applied such that these effects are minimised (e.g. driving it through the pair which is at maximal distance from both).

*4.2 Study technical limitations and future work*

First, the study was limited to only consider a bipolar current source and a bi-phasic symmetric current pulse. It was done on the basis of engineering practicality and reliability; however, authors admit that further investigations are required, and a 2-electrode longitudinal configuration might not be the best in the case of a non-symmetric current pulse.

Secondly, the study was only limited to a particular myelinated fibre. This was done with the applicability to the existing trends in implantable stimulators and the particular diseases; however, fibres of other diameters and non-myelinated fibres are currently under investigation. It is expected that the fibre excitation threshold would change as the fibre diameter changes. So although it may not significantly alter the results when all of the nerve fibres would be scaled, it will undoubtfully have an effect when fibres are considered within a mixed nerve. Fibre diameter would also change the spatial distribution of the nodes of Ranvier along the fibre length, so it might have a more significant effect on the activation function and would lead to a different optimal distance between the electrodes. Future work will include a histology-based approximation of the fibres distribution and location.

Thirdly, the study was limited to a generic nerve model not accounting for fascicular structure and, in particular, anisotropic properties of the nerve. Previous studies [31], however, suggest that this does not affect the applicability of the model for optimisation purposes without *a priori* knowledge about the fascicular structure.

The optimisation strategy in this paper was to target the specific fibres within the nerve while avoiding stimulation of all other fibres by minimising the affected area. The alternative approach would be to target a specific area on the fascicular level, whilst minimising the activation of all other fascicles. This approach could lead to slightly different results, but for this one would need to know the spatial fascicular organisation of the nerve in question, and the mismatch between the geometries could produce larger errors.

In this study, we aimed to show the full extent of variation of the OCF over the parameter space so the approach which was used for the optimisation involved the scanning of the entire parameter space. This is not the best method for the optimisation, and actual optimisation methods could potentially speed up the process and result in a more precisely optimised device. Some of the geometric parameters of the model, e.g. the fact that electrodes were rectangular etc., were constrained because of the used optimisation method. Although it is unlikely that the above geometrical constraints could significantly alter the optimal configuration in our particular case of the biphasic symmetric current pulse, however, it could play a significant role in more complicated stimulation paradigms. In this case, the use of more generic optimisation approaches like bayesian or evolutionary methods would be preferable as the simple grid search would not be up to the task.

Finally, the use of sheep as an animal model allowed using human-size vagus nerves as well as a model very often used for studying cardiac functions. However, in the sheep, the nerve may contain fibres innervating the rumen, which is not human-related. It is therefore planned to validate the array in other large animal species such as a pig.

Future work will be concentrated on the further optimisation of stimulation using multiple current sources, as well as fibre-specific and direction-specific stimulation paradigms, with the view of delivering the ultimate device for spatial, directional and fibre-selective therapeutic neuromodulation in complex nerves.

**Acknowledgements**

This research was supported by GSK-UCL collaboration grant "Imaging and selective stimulation of autonomic nerve traffic using Electrical Impedance Tomography and a non-penetrating nerve cuff".

Authors would like to thank Professor Nick Donaldson and Implantable Devices Group (IDG) at UCL for providing the cleanroom for electrodes manufacturing.

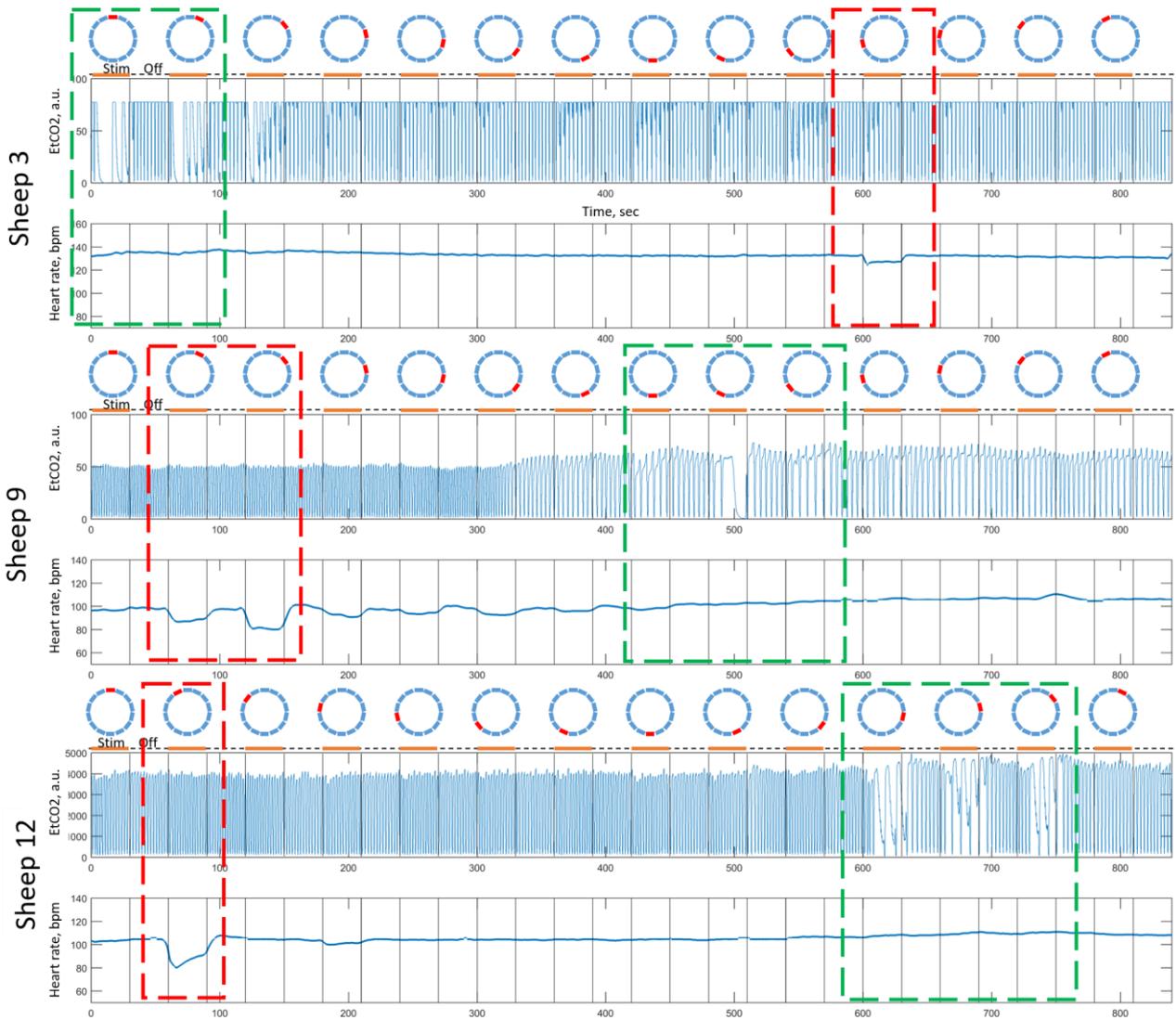

Figure 9. Three examples of the experimental results at optimal current levels. The stimulus pulse was delivered to the longitudinal electrode pair at each of the 14 possible circumferential "o'clock" locations (indicated by the pictograms above) for 30 s, followed by a 30 s interstimulation interval. In all animals, there were pairs which induced significant alterations in breathing pattern (down to full apnoea, green squares), whilst others caused significant bradycardia (red square).

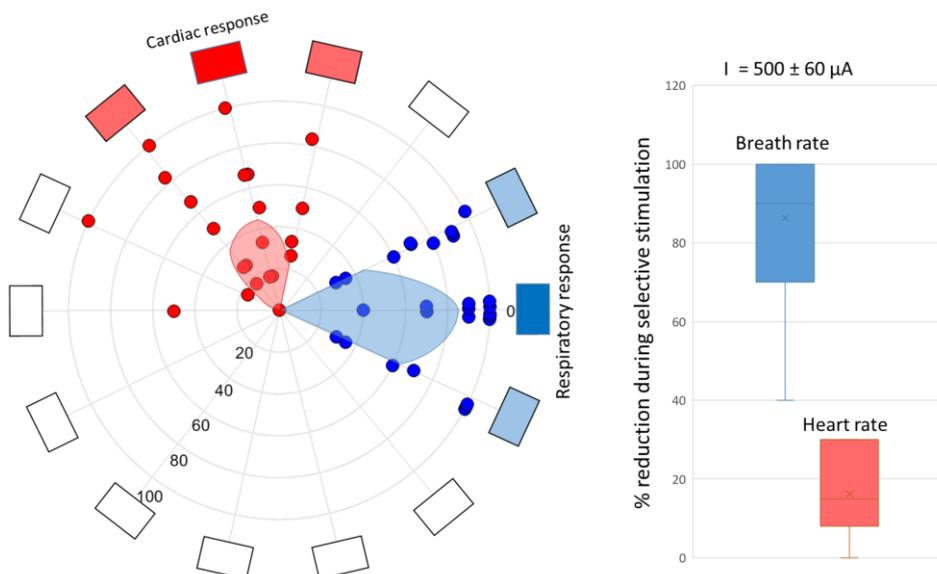

Figure 10. Summary of results across all animals. Each dot represents the significant respiration- (blue dots) and cardiac- (red dots) specific "o'clock" site, and aligned by most respiration-specific. The radial position indicates the normalised % change from the baseline (maximum 100% for RBR and 32% for HR, shown on the right, N = 24 in 12 sheep).

**Competing interests**

M.D and D.C. are employees of Galvani Bioelectronics. Authors declare that University College London has filed international patent applications describing methods of selective nerve modulation. Authors declare that Galvani Bioelectronics provided funds to support the work associated with this manuscript.

**References**

[1] K. Famm, B. Litt, K. J. Tracey, E. S. Boyden, and M. Slaoui, "Drug discovery: a jump-start for electroceuticals.," *Nature*, vol. 496, no. 7444, pp. 159–61, Apr. 2013.

[2] K. Birmingham *et al.*, "Bioelectronic medicines: a research roadmap.," *Nat. Rev. Drug Discov.*, vol. 13, no. 6, pp. 399–400, Jun. 2014.

[3] P. S. Olofsson and K. J. Tracey, "Bioelectronic medicine: technology targeting molecular mechanisms for therapy.," *J. Intern. Med.*, vol. 282, no. 1, pp. 3–4, Jul. 2017.

[4] R. S. McLachlan, "Vagus nerve stimulation for intractable epilepsy: a review.," *J. Clin. Neurophysiol.*, vol. 14, no. 5, pp. 358–68, Sep. 1997.

[5] M. Li, T. Sato, T. Kawada, M. Sugimachi, and K. Sunagawa, "Vagal Nerve Stimulation Markedly Improves Long-Term Survival After Chronic Heart Failure in Rats," *Circulation*, vol. 109, pp. 120–124, 2004.

[6] H. N. Sabbah, I. Ilsar, A. Zaretsky, S. Rastogi, M. Wang, and R. C. Gupta, "Vagus nerve stimulation in experimental heart failure," *Heart Fail. Rev.*, vol. 16, no. 2, pp. 171–178, Mar. 2011.

[7] E. C. Meyers *et al.*, "Vagus Nerve Stimulation Enhances Stable Plasticity and Generalization of Stroke Recovery," *Stroke*, vol. 49, no. 3, pp. 710–717, Mar. 2018.

[8] F. A. Koopman *et al.*, "Vagus nerve stimulation inhibits cytokine production and attenuates disease severity in rheumatoid arthritis.," *Proc. Natl. Acad. Sci. U. S. A.*, vol. 113, no. 29, pp. 8284–9, Jul. 2016.

[9] A. Mertens, R. Raedt, S. Gadeyne, E. Carrette, P. Boon, and K. Vonck, "Recent advances in devices for vagus nerve stimulation," *Expert Rev. Med. Devices*, vol. 15, no. 8, pp. 527–539, Aug. 2018.

[10] D. Guiraud *et al.*, "Vagus nerve stimulation: state of the art of stimulation and recording strategies to address autonomic function neuromodulation," *J. Neural Eng.*, vol. 13, no. 4, p. 041002, Aug. 2016.

[11] T. Sukhu, M. J. Kennelly, and R. Kurpad, "Sacral neuromodulation in overactive bladder: a review and current perspectives.," *Res. reports Urol.*, vol. 8, pp. 193–199, 2016.

[12] L. L. de Wall and J. P. Heesakkers, "Effectiveness of percutaneous tibial nerve stimulation in the treatment of overactive bladder syndrome.," *Res. reports Urol.*, vol. 9, pp. 145–157, 2017.

[13] K. Dunning, M. W. O'Dell, P. Kluding, and K. McBride, "Peroneal Stimulation for Foot Drop After Stroke," *Am. J. Phys. Med. Rehabil.*, vol. 94, no. 8, pp. 649–664, Aug. 2015.

[14] S. O. Hong, Y.-F. Chen, J. Jung, Y.-D. Kwon, and S. Y. C. Liu, "Hypoglossal nerve stimulation for treatment of obstructive sleep apnea (OSA): a primer for oral and maxillofacial surgeons.," *Maxillofac. Plast. Reconstr. Surg.*, vol. 39, no. 1, p. 27, Dec. 2017.

[15] D. W. Dodick *et al.*, "Safety and efficacy of peripheral nerve stimulation of the occipital nerves for the management of chronic migraine: Long-term results from a randomized, multicenter, double-blinded, controlled study.," *Cephalalgia*, pp. 0333102414543331-, Jul. 2014.

[16] E. Ben-Menachem, "Vagus nerve stimulation, side effects, and long-term safety.," *J. Clin. Neurophysiol.*, vol. 18, no. 5, pp. 415–8, Sep. 2001.

[17] D. J. Tyler and D. M. Durand, "Functionally selective peripheral nerve stimulation with a flat interface nerve electrode," *IEEE Trans. Neural Syst. Rehabil. Eng.*, vol. 10, no. 4, pp. 294–303, Dec. 2002.

[18] B. Wodlinger and D. M. Durand, "Selective recovery of fascicular activity in peripheral nerves," *J. Neural Eng.*, vol. 8, no. 5, p. 056005, Oct. 2011.

[19] E. L. Graczyk, M. A. Schiefer, H. P. Saal, B. P. Delhaye, S. J. Bensmaia, and D. J. Tyler, "The neural basis of perceived intensity in natural and artificial touch.," *Sci. Transl. Med.*, vol. 8, no. 362, p. 362ra142, Oct. 2016.

[20] K. S. Mathews, H. A. C. Wark, and R. A. Normann, "Assessment of rat sciatic nerve function following acute implantation of high density utah slanted electrode array (25 electrodes/mm2) based on neural recordings and evoked muscle activity," *Muscle and Nerve*, vol. 50, no. 3, pp. 417–424, 2014.

[21] G. A. Clark, N. M. Ledbetter, D. J. Warren, and R. R. Harrison, "Recording sensory and motor information from peripheral nerves with Utah

Slanted Electrode Arrays," in *2011 Annual International Conference of the IEEE Engineering in Medicine and Biology Society*, 2011, vol. 2011, pp. 4641–4644.

[22] H. A. C. Wark, K. S. Mathews, R. A. Normann, and E. Fernandez, "Behavioral and cellular consequences of high-electrode count Utah Arrays chronically implanted in rat sciatic nerve," *J Neural Eng*, vol. 11, p. 046027, 2014.

[23] M. B. Christensen, H. A. C. Wark, and D. T. Hutchinson, "A histological analysis of human median and ulnar nerves following implantation of Utah slanted electrode arrays," *Biomaterials*, vol. 77, pp. 235–242, Jan. 2016.

[24] D. T. T. Plachta et al., "Blood pressure control with selective vagal nerve stimulation and minimal side effects," *J. Neural Eng.*, vol. 11, no. 3, p. 036011, Jun. 2014.

[25] J. Rozman and M. Bunc, "Modulation of visceral function by selective stimulation of the left vagus nerve in dogs," *Exp. Physiol.*, vol. 89, no. 6, pp. 717–725, Nov. 2004.

[26] S. C. M. A. Ordelman, L. Kornet, R. Cornelussen, H. P. J. Buschman, and P. H. Veltink, "Selectivity for specific cardiovascular effects of vagal nerve stimulation with a multi-contact electrode cuff.," *IEEE Trans. Neural Syst. Rehabil. Eng.*, vol. 21, no. 1, pp. 32–6, Jan. 2013.

[27] P. Pečlin et al., "Selective stimulation of the vagus nerve in a man," Springer, Berlin, Heidelberg, 2009, pp. 1628–1631.

[28] M. Lotfi Navaii, H. Sadjedi, and M. Jalali, "Waveform efficiency analysis of auditory nerve fiber stimulation for cochlear implants.," *Australas. Phys. Eng. Sci. Med.*, vol. 36, no. 3, pp. 289–300, Sep. 2013.

[29] A. Wongsarnpigoon, J. P. Woock, and W. M. Grill, "Efficiency Analysis of Waveform Shape for Electrical Excitation of Nerve Fibers," *IEEE Trans. Neural Syst. Rehabil. Eng.*, vol. 18, no. 3, pp. 319–328, Jun. 2010.

[30] W. M. Grill, S. E. Norman, and R. V. Bellamkonda, "Implanted Neural Interfaces: Biochallenges and Engineered Solutions," *Annu. Rev. Biomed. Eng.*, vol. 11, no. 1, pp. 1–24, Aug. 2009.

[31] M. Dali et al., "Model based optimal multipolar stimulation without *a priori* knowledge of nerve structure: application to vagus nerve stimulation," *J. Neural Eng.*, vol. 15, no. 4, p. 046018, Aug. 2018.

[32] A. R. Kent and W. M. Grill, "Model-based analysis and design of nerve cuff electrodes for restoring bladder function by selective stimulation of the pudendal nerve.," *J. Neural Eng.*, vol. 10, no. 3, p. 036010, Jun. 2013.

[33] A. Q. Choi, J. K. Cavanaugh, and D. M. Durand, "Selectivity of multiple-contact nerve cuff electrodes: a simulation analysis," *IEEE Trans. Biomed. Eng.*, vol. 48, no. 2, pp. 165–172, 2001.

[34] E. N. Warman, W. M. Grill, and D. Durand, "Modeling the effects of electric fields on nerve fibers: Determination of excitation thresholds," *IEEE Trans. Biomed. Eng.*, vol. 39, no. 12, pp. 1244–1254, 1992.

[35] D. Calvetti, B. Wodlinger, D. M. Durand, and E. Somersalo, "Hierarchical beamformer and cross-talk reduction in electroneurography," *J. Neural Eng.*, vol. 8, no. 5, p. 056002, Oct. 2011.

[36] E. Somersalo, M. Cheney, and D. Isaacson, "Existence and Uniqueness for Electrode Models for Electric Current Computed Tomography," *SIAM J. Appl. Math.*, vol. 52, no. 4, pp. 1023–1040, Aug. 1992.

[37] C. A. R. Chapman et al., "Electrode fabrication and interface optimization for imaging of evoked peripheral nervous system activity with electrical impedance tomography (EIT)," *J. Neural Eng.*, vol. 16, no. 1, p. 016001, Feb. 2019.

[38] F. Rattay, "Analysis of models for extracellular fiber stimulation," *IEEE Trans. Biomed. Eng.*, vol. 36, no. 7, pp. 676–682, Jul. 1989.

[39] I. Tarotin, K. Aristovich, and D. Holder, "Simulation of impedance changes with a FEM model of a myelinated nerve fibre," *J. Neural Eng.*, vol. 16, no. 5, p. 056026, Sep. 2019.

[40] J. Howells, L. Trevillion, H. Bostock, and D. Burke, "The voltage dependence of I h in human myelinated axons," *J. Physiol.*, vol. 590, no. 7, pp. 1625–1640, 2012.

[41] C.-H. Berthold and M. Rydmark, "Anatomy of the paranode-node-paranode region in the cat," *Experientia*, vol. 39, no. 9, pp. 964–976, Sep. 1983.

[42] C. C. McIntyre, A. G. Richardson, and W. M. Grill, "Modeling the excitability of mammalian nerve fibers: influence of afterpotentials on the recovery cycle.," *J. Neurophysiol.*, vol. 87, no. 2, pp. 995–1006, 2002.

[43] I. Tarotin, "COMSOL model of a myelinated fibre

bi-directionally coupled with external space," 2019. [Online]. Available: https://github.com/EIT-team/Myelinated-fibre-model.

[44] M. L. Settell *et al.*, "Functional vagotopy in the cervical vagus nerve of the domestic pig: Implications for the study of vagus nerve stimulation," *J. Neural Eng.*, 2020.

[45] U. Ahmed *et al.*, "Anodal block permits directional vagus nerve stimulation," *Sci. Rep.*, 2020.

[46] E. R. Spitzer and M. L. Hughes, "Effect of stimulus polarity on physiological spread of excitation in cochlear implants," *J. Am. Acad. Audiol.*, 2017.

[47] O. Macherey, R. P. Carlyon, A. Van Wieringen, J. M. Deeks, and J. Wouters, "Higher sensitivity of human auditory nerve fibers to positive electrical currents," *JARO - J. Assoc. Res. Otolaryngol.*, 2008.